\documentclass[amssymb,superscriptaddress,twocolumn,prl]{revtex4}
\usepackage{graphicx}
\def \be{\begin{equation}}
\def \ee{\end{equation}}
\def \bea{\begin{eqnarray}}
\def \eea{\end{eqnarray}}

\newcommand{\beq}[1]{\begin{eqnarray}\label{#1}}
\newcommand{\eeq}{\end{eqnarray}}

\begin{document}
 \pagestyle{plain}

 \title{Spherical Collapse Model}

\author{Ding-fang Zeng and Yi-hong Gao}
\email{dfzeng@itp.ac.cn} \email{gaoyh@itp.ac.cn}
\affiliation{Institute of Theoretical Physics, Chinese Academy of
Science.}

 \begin{abstract}
 We studied the spherical collapse model in the
 flat $\Lambda$CDM cosmology and provided exact and analytical formulaes for
 the calculation of the two important
 parameters in the application of Press-Schechter theory, the
 critical density contrast and the ratio of cluster/background
 densities at the virialization point in terms of
 a media variable which can be solved precisely by numerical methods.
 \end{abstract}

%\pacs{04.60.Pp, 04.60.Kz, 98.80.Jk}

 \maketitle

 \section{Introduction}

 Press-Schechter theory has been written into textbooks, but two
 important parameters $\delta_c$ and $\Delta_c$ of it extracted from the studying of
 spherical collapse models are not determined well enough. \cite{GunnGott, LaceyCole}
 are two early works studying the spherical models. They considered
 this model in a background universe without dark energies.
 \cite{Barrow, ECF, WangSteinhardt1, LokasHoffman} are four later works studying this
 model in the flat $\Lambda$CDM or open cosmologies.
 Our purpose here is to provide a simple and direct method
 for the calculation of
 $\delta_c(\Omega_{m0},a_c)$ and $\Delta_c(\Omega_{m0},a_c)$. Our
 background is flat $\Lambda$CDM cosmology but the method can be easily
 generalized to the QCDM \cite{WangSteinhardt1} as well as
 non-flat cosmologies. Comparing with the existed results on this
 model, our results are analytical and exact.

 \section{Press-Schechter Theory}

 Press-Schechter Theory \cite{PressSchechter} predicts that the
 fraction of volume which has collapsed at a certain red-shift $z$ is
 \beq{}
 f_{coll}(M(R),z)=\frac{2}{\sqrt{2\pi}\sigma(R,z)}
 \int_{\delta_c}^{\infty}d\delta e^{-\delta^2/2\sigma^2(R,z)}.
 \nonumber\\
 \label{Press-Schechter}
 \eeq
 Here, $R$ is the radius over which the density field has been
 smoothed, $\sigma(R,z)$ is
 the rms of the smoothed density
 field \cite{Dodelson}. $\delta_c$ is the threshold of density contrast at time
 $z$ (count stopping time)
 beyond which objects collapse.

 In existing literatures, there are two conventions for the
 definition of $\sigma(R,z)$ and $\delta_c$. In the first
 convention \cite{Dodelson, WangSteinhardt1, VianaLiddle},
 $\sigma(R,z)$ contains growth factor,
 while $\delta_c$ only depends on the partition of
 the cosmological component today and the count stopping epoch $a_c$.
 In the second convention \cite{ECF}, $\sigma(R)$ does not contain the growth factor,
 but $\delta_c$ contains both the cosmological component partition
 effects and the growth factor effects. We will take the first
 convention in this paper. Our $\delta_c$ here will be
 corresponded to the $\delta_c(0)$ of \cite{ECF}, which was
 plotted in the upper panel of figure.1 of it.

 Operationally, $\delta_c$ is obtained by extrapolating the primordial
 perturbation to the collapse epoch using the growth law of
 linear perturbation theory, i. e.,
 \beq{}
 \delta_c=\left[(\frac{\rho_{mc}(a,a_c)}{\rho_{mb}(a)}-1)\frac{1}{D_1(a)}\right]_{a\rightarrow0}D_1(a_c)
 \label{deltacDefinition}
 \eeq
 Where $\rho_{mc}$ and $\rho_{mb}$ are the mass density of the
 clusters and the background respectively; while $D_1(a)$ is the
 growth function of linear perturbation theory \cite{Dodelson},
 \beq{}
 D_1(a)=\frac{5\Omega_{m0}H^2_0}{2}H(a)\int_0^ada^{\prime}[a^{\prime}H(a^{\prime})]^{-3}.
 \eeq
 It can be shown that $D_1(a)_{a\rightarrow0}\rightarrow a$. Using the method of
 \cite{LaceyCole} we can show that
 \beq{}
 \left[\frac{r}{a}\right]_{a\rightarrow0}\propto(1-\alpha\cdot a)
 \label{roaagoesto0}
 \eeq
 So,
 \beq{}
 \delta_c=3\alpha\cdot D_1(a_c).
 \label{deltacOperationDefinition}
 \eeq
 In a given cosmological model, $D_1(a)$ are known
 \cite{Dodelson}, the purpose of studying spherical collapse
 model is to determine $\alpha$.

 It should be noted that besides the partition of the
 cosmological components and the observational epoch, the value
 of $\delta_c$ is also quite dependent on the choice of
 smoothing window used to obtain the dispersion
 $\sigma(R,z)$ \cite{LaceyCole}. We will not consider
 this effect in this paper.

 \section{Spherical Collapse Model}

 To calculate parameter $\alpha$ of eq(\ref{deltacOperationDefinition}),
 let us start with the Friedman equations for
 both the over-dense region and back ground cosmology:
 \beq{}
 (\frac{\dot{r}}{r})^2&&\hspace{-3mm}=\frac{8\pi G}{3}[\frac{\rho_{mc,ta}r^3_{ta}}{r^3}+\rho_{\Lambda 0}]-\frac{\kappa}{r^2}
 \label{FriedmanC} \\
 (\frac{\dot{a}}{a})^2&&\hspace{-3mm}=\frac{8\pi G}{3}[\frac{\rho_{mb,ta}a^3_{ta}}{a^3}+\rho_{\Lambda 0}]
 \label{FriedmanB}
 \eeq
 , where $r$ and $a$ denote the radius of the over-dense region
 and scale factor of background universe respectively; $\kappa$ is
 a constant; while subscript $_{ta}$ and $_c$ denote
 the turn around and collapse point time. Because at the turn around time $\dot{r}=0$,
 \beq{}
 \frac{\kappa}{r^2_{ta}}=
 \frac{8\pi G}{3}[\rho_{mc,ta}+\rho_{\Lambda 0}].
 \label{kappa}
 \eeq
 Dividing eq(\ref{FriedmanC}) by (\ref{FriedmanB}) and using (\ref{kappa}) we get
 \beq{}
 \frac{\dot{r}^2r^{-2}_{ta}}{\dot{a}^2a^{-2}_{ta}}
 =\frac{\rho_{mc,ta}r^{-1}r_{ta}+\rho_{\Lambda 0}r^2r^{-2}_{ta}-(\rho_{mc,ta}+\rho_{\Lambda 0})}{\rho_{mb,ta}a^{-1}a_{ta}+\rho_{\Lambda
 0}a^2a^{-2}_{ta}}
 \label{FriedmannV}
 \eeq
 After letting
 \beq{}
 x=\frac{a}{a_{ta}},y=\frac{r}{r_{ta}},\zeta=\frac{\rho_{mc,ta}}{\rho_{mb,ta}},\nu=\frac{\rho_{\Lambda0}}{\rho_{mb,ta}}
 \label{x-y-zeta-nu-definition}
 \eeq
 eq(\ref{FriedmannV}) becomes
 \beq{}
 (\frac{dy}{dx})^2=\frac{\zeta y^{-1}+\nu y^2-(\zeta+\nu)}{x^{-1}+\nu
 x^2}.
 \label{dydxMainEq}
 \eeq
 We will not try to solve this equation, we derive it here
 because we need to use its asymptotic behavior to calculate
 $\delta_c$.
 %This equation is a little different from eq(A1) of
 %\cite{ECF}. There the coefficients of second power term in both
 %the denominator and numerator are the same, while here they are
 %not. This difference can be cancelled by redefinition of $y$.

 In the $\Lambda$CDM cosmologies, the quantity $\zeta$ and
 $\nu$ in eq(\ref{dydxMainEq}) are only functions of $\Omega_{m0}$ and $a_c$.
 They do not depends on $x$ or $y$. Using eq(\ref{roaagoesto0}),
 we can write
 \beq{}
 \left[\frac{y}{x}\right]_{x\rightarrow0}=
 \left[\frac{rr^{-1}_{ta}}{aa^{-1}_{ta}}\right]_{a\rightarrow0}=\zeta^{\frac{1}{3}}(1-\alpha\cdot
 a_{ta}x).
 \label{yoverx}
 \eeq
 So,
 \beq{}
 \frac{d}{dx}[y_{x\rightarrow0}]=\zeta^{\frac{1}{3}}(1-2\alpha\cdot a_{ta}x).
 \label{dydx}
 \eeq
 Substitute eqs(\ref{yoverx}) and (\ref{dydx}) into
 eq(\ref{dydxMainEq}), expand it and keep only the linear
 term, we get
 \beq{}
 \alpha=\frac{1}{5}a^{-1}_{ta}[\zeta^{\frac{1}{3}}+\nu\cdot
 a_{ta}\zeta^{-\frac{2}{3}}].
 \label{alpha-result}
 \eeq
 Substitute this result into eq(\ref{deltacOperationDefinition}),
 we get:
 \beq{}
 \delta_c(\Omega_{m0},a_c)=\frac{3}{5}a^{-1}_{ta}[\zeta^{\frac{1}{3}}+\nu\cdot
 a_{ta}\zeta^{-\frac{2}{3}}]D_1(a_c)
 \label{deltacAnalytical}
 \eeq
 This is the first important result of this paper. We will
 derive the differential equation for determining $\zeta(\Omega_{m0},a_c)$ and
 the fitting formula eq(\ref{zetaLCDMold}) for it as well as the analytical solution for
 $\nu(\Omega_{m0},a_c)$ eq(\ref{LCDMtcAndtta}) (and the notes there) in the technique details
 section. Comparing with the method of \cite{ECF, WangSteinhardt1, LokasHoffman}
 our method here is more easier to operate and the result is more
 simply-looking.

 \begin{figure}
 \includegraphics[scale=0.4]{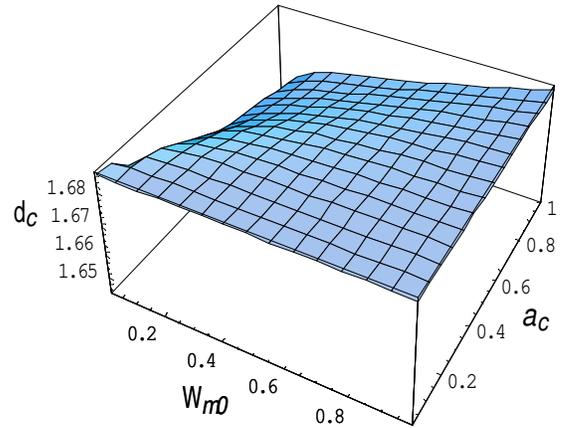}
 \caption{$\delta_c^{\prime}$s
 dependence on the cosmological models characterized by
 $\Omega_{m0}$ and the collapse epoch $a_c$.
 }\label{deltac3D}
 \end{figure}
 \begin{figure}
 \includegraphics[scale=0.4]{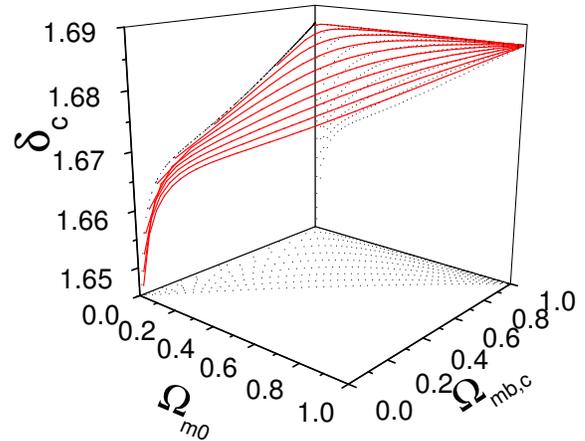}
 \caption{
 $\delta_c^{\prime}$s
 dependence on $\Omega_{m0}$ and
 the collapse epoch matter percentage $\Omega_{mb,c}$.
 Points on the same space curves have the same
 collapse epoch $a_c$. The outmost curve has $a_c=1$,
 the next one has $a_c=0.9$ and so on. The deeper a
 curve lies in the paper, the smaller $a_c$ it has.
 }\label{ECFd}
 \end{figure}

 In FIG.\ref{deltac3D} we plotted
 $\delta_c^{\prime}$s dependence on $\Omega_{m0}$ and $a_c$.
 From the figure we see that if $\Omega_{m0}$ is fixed, then
 $\delta_c$ is a decreasing function of $a_c$. Physically this is
 because, in the given cosmological model, $\Omega_{m0}$ fixed,
 the earlier an objects collapse, the more denser it is required
 to assure collapse occur successfully. On the other hand, if
 $a_c$ is fixed, then $\delta_c$ increases as $\Omega_{m0}$
 increases. Superficially this means that in those lower matter (thus high dark energy)
 percentage universes, collapse occurs more easily! This is not
 the fact. Because in the lower matter percentage universes,
 density perturbation grows more slowly. As the result of linear
 extrapolation of the primary perturbations
 eq(\ref{deltacDefinition}), the more smaller growth factor comes from
 the lower matter percentage suppresses the primary perturbations
 more strongly so gives more smaller $\delta_c$.

 To compare our results with that of \cite{ECF}, we plotted
 $\delta_c^\prime$s dependence on $\Omega_{m0}$ and
 $\Omega_{mb,c}$ in FIG.\ref{ECFd}. From this figure we see that
 what was shown in the upper panel of figure.1 of \cite{ECF} is
 the projection of our results in the $\delta_c$-$\Omega_{mb,c}$
 plane.

 To this point in this section, we imagined the evolution of an
 over-dense region as: as long as it is over-dense enough,
 it will grow from an $r=0$ point, to the maximum
 radius, then collapse to $r=0$ point. Factually, before the
 region collapse to the $r=0$ point, pressures from the random
 moving of the materials inside the region will balance their
 self-gravity and the system will enter virialization status.
 At this point, the second important parameter
 $\Delta_c:=\frac{\rho_{mc,vir}}{\rho_{mb,vir}}$ in the application of
 Press-Schechter theory \cite{ECF, VianaLiddle, WangSteinhardt1} appears.
 Assuming that at the collapse point, the system has virialized fully, we can write:
 \beq{}
 \Delta_c=
 \frac{\rho_{mc}(a_{ta})r^3_{ta}r^{-3}_c}{\rho_{mb}(a_{ta})a^3_{ta}a^{-3}_c}
 =\zeta\frac{r^3_{ta}}{r^3_c}\frac{a_c^3}{a_{ta}^3}.
 \label{DeltacDefinition}
 \eeq
 According to virial theorem and energy conservation law \cite{WangSteinhardt1},
 \beq{}
 &&\hspace{-5mm}E_{kinetic}=-\frac{1}{2}U_{G}+U_{\Lambda}\nonumber\\
 &&\hspace{-5mm}\frac{1}{2}U_{G,c}+2U_{\Lambda,c}=U_{G,ta}+U_{\Lambda,ta}
 \eeq
 we have
 \beq{}
 &&\hspace{-5mm}-\frac{1}{2}\frac{3GM^2}{5r_c}-2\cdot\frac{4\pi GM\rho_{\Lambda,c} r^2_c}{5}
 \nonumber\\
 &&\hspace{18mm}=-\frac{3GM^2}{5r_{ta}}-\frac{4\pi GM\rho_{\Lambda,ta} r^2_{ta}}{5}
 \nonumber\\
 &&\Rightarrow\nonumber\\
 &&\hspace{-5mm}\frac{r_{ta}}{r_c}=
 \frac{2(\rho_{mc,ta}+\rho_{\Lambda,ta})\rho^{-1}_{mc,ta}}{(\rho_{mc,c}+4\rho_{\Lambda,c})\rho^{-1}_{mc,c}}
 \nonumber\\
 &&\hspace{5mm}=\frac{2(1+\frac{\Omega_{\Lambda,ta}}{\Omega_{mb,ta}}\frac{1}{\zeta})}
 {1+4\frac{\Omega_{\Lambda,c}}{\Omega_{mb,c}}[\zeta\frac{r^3_{ta}}{r^3_c}\frac{a_c^3}{a^3_{ta}}]^{-1}}
 \label{RtaORc}
 \eeq
 Looking as an equation for $\frac{r_{ta}}{r_c}$, eq(\ref{RtaORc})
 can be solved analytically. Substituting the solution into
 eq(\ref{DeltacDefinition}), we get
 \beq{}
 &&\hspace{-3mm}\Delta_c(\Omega_{m0},a_c)=\zeta\cdot\left[\frac{a_c}{a_{ta}}\right]^3\nonumber\\
 &&\hspace{3mm}\times\left\{\frac{2}{3}[\frac{(\mu_t+1)^2}{f(\mu_t,\mu_c)}+(\mu_t+1)+f(\mu_t,\mu_c)]\right\}^{3}
 \label{NLDeltacAnalytical}\\
 &&\hspace{-3mm}f(\mu_t,\mu_c)=[(\mu_t+1)^3-\frac{27}{4}\mu_c
 \nonumber\\
 &&\hspace{15mm}+\frac{\sqrt{27}}{4}\sqrt{\mu_c(27\mu_c-8(\mu_t+1)^3)}]^{1/3}
 \label{fdefinition}\\
 &&\mu_t=\frac{\Omega_{\Lambda,ta}}{\Omega_{mb,ta}}\frac{1}{\zeta},\
 \mu_c=\frac{\Omega_{\Lambda,c}}{\Omega_{mb,c}}\frac{1}{\zeta}\frac{a^3_{ta}}{a^3_c}
 \label{mutmucdefinition}
 \eeq
 Eq(\ref{NLDeltacAnalytical}) is the second important result of this
 paper. In $\Lambda$CDM cosmology, all the quantities on the right
 hand side of eq(\ref{NLDeltacAnalytical}) are known functions of
 $\Omega_{m0}$ and $a_c$. We will give the relevant formulaes
 eq(\ref{OmegaLCDM}), (\ref{zetaLCDMold}) and (\ref{LCDMtcAndtta})
 in the technique section. Comparing with \cite{WangSteinhardt1},
 our results here is exact instead of approximated.

\begin{figure}
 \includegraphics[scale=0.22]{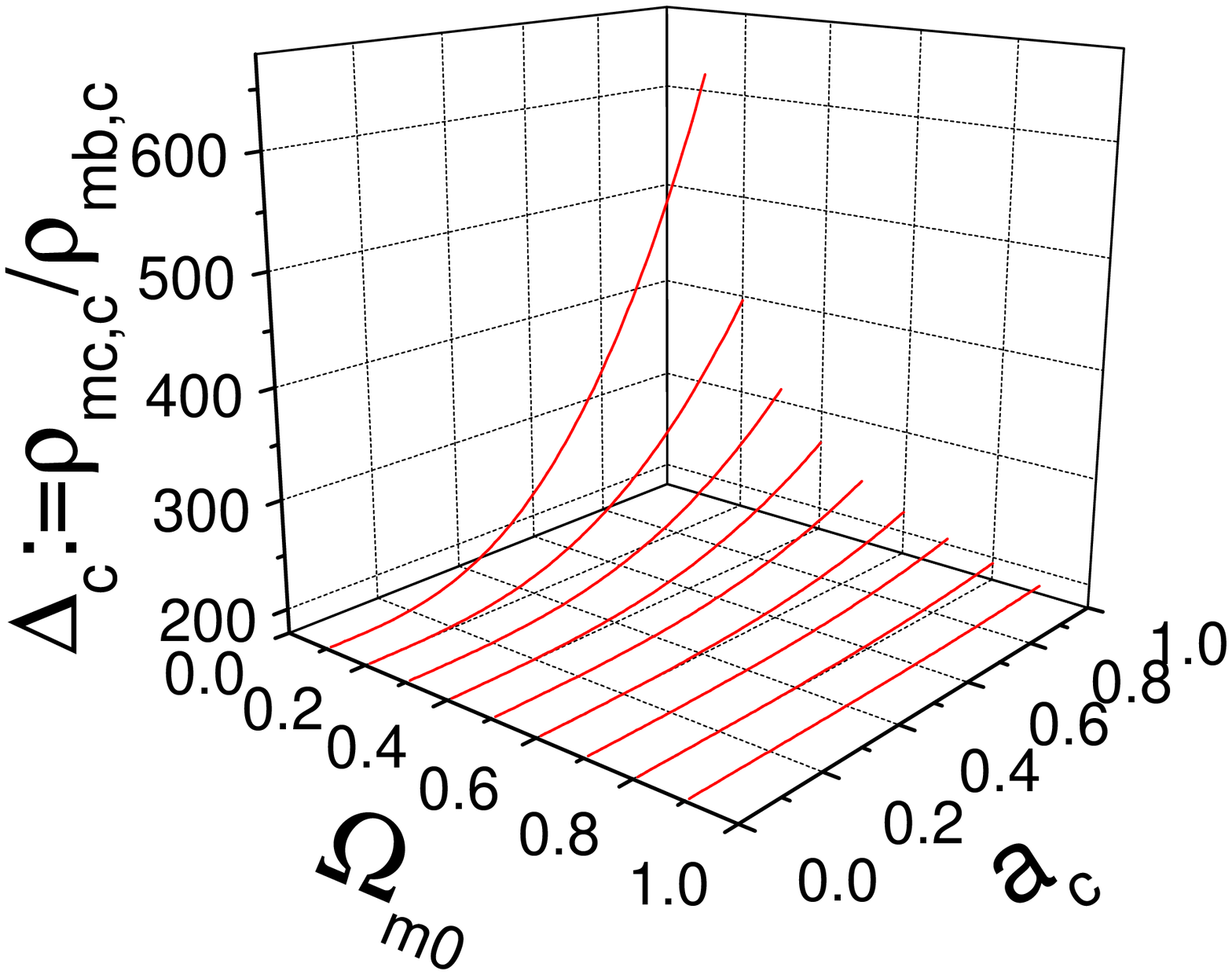}
 \includegraphics[scale=0.22]{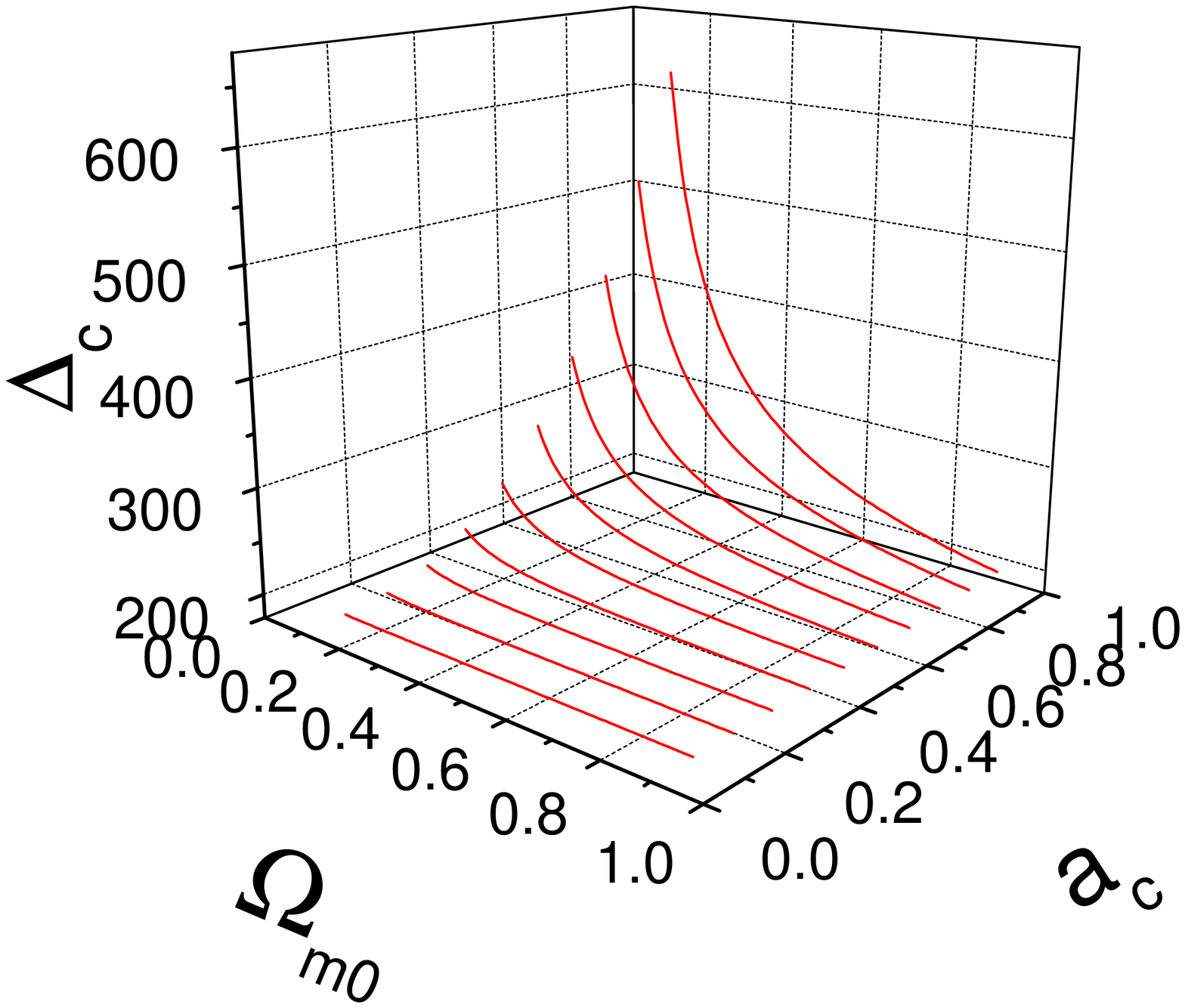}
 \caption{Left panel: $\Delta_c^{\prime}$s
 dependence on the collapse epoch $a_c$, different
 curves denote different cosmological models indexed by $\Omega_{m0}$.
 Right panel: $\Delta_c^{\prime}$s
 dependence on the cosmological models, different
 curves denote different collapse epoch indexed by $a_c$.
 }\label{NLD3D}
 \end{figure}

\begin{figure}
 \includegraphics[scale=0.22]{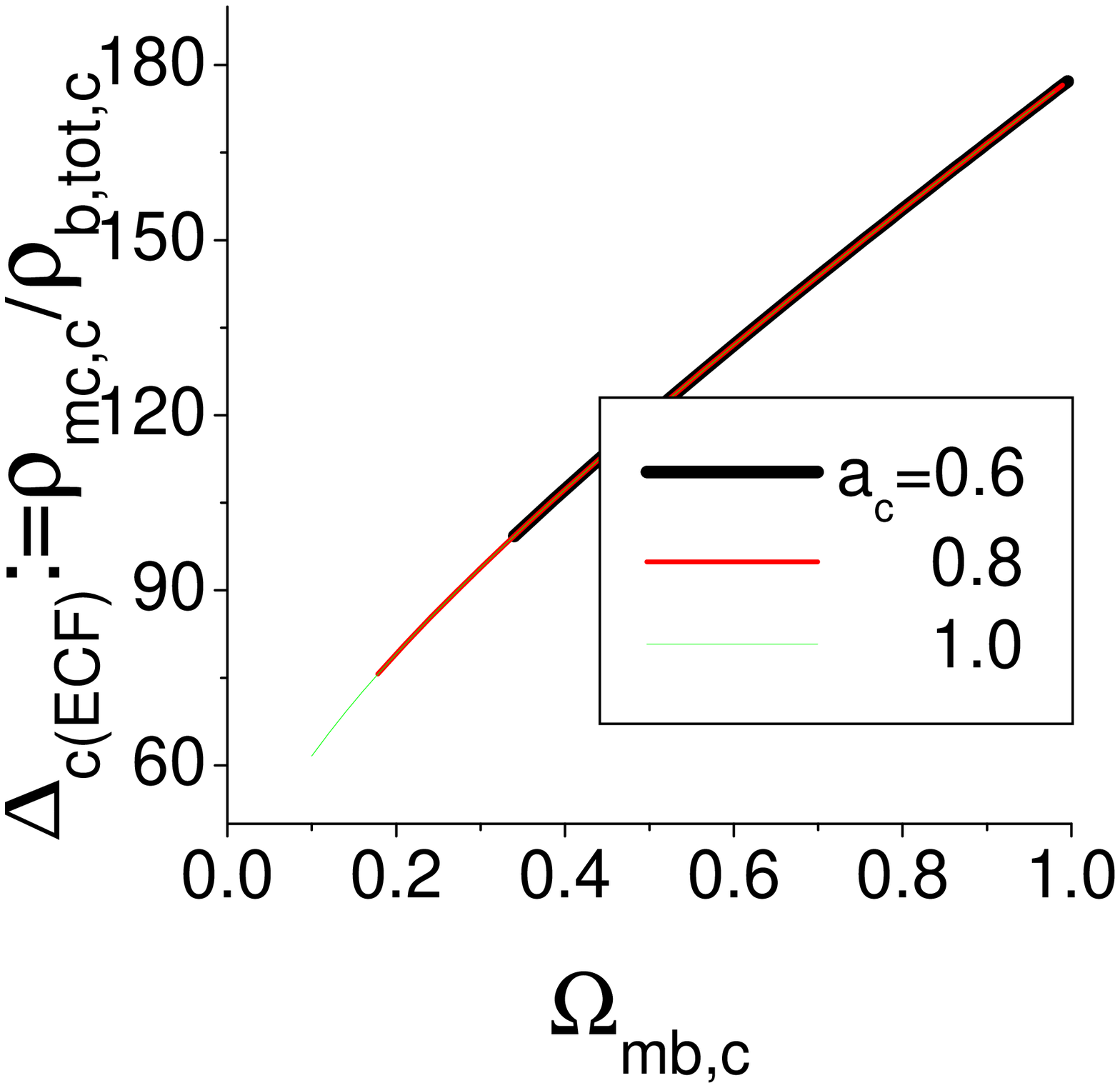}
 \includegraphics[scale=0.22]{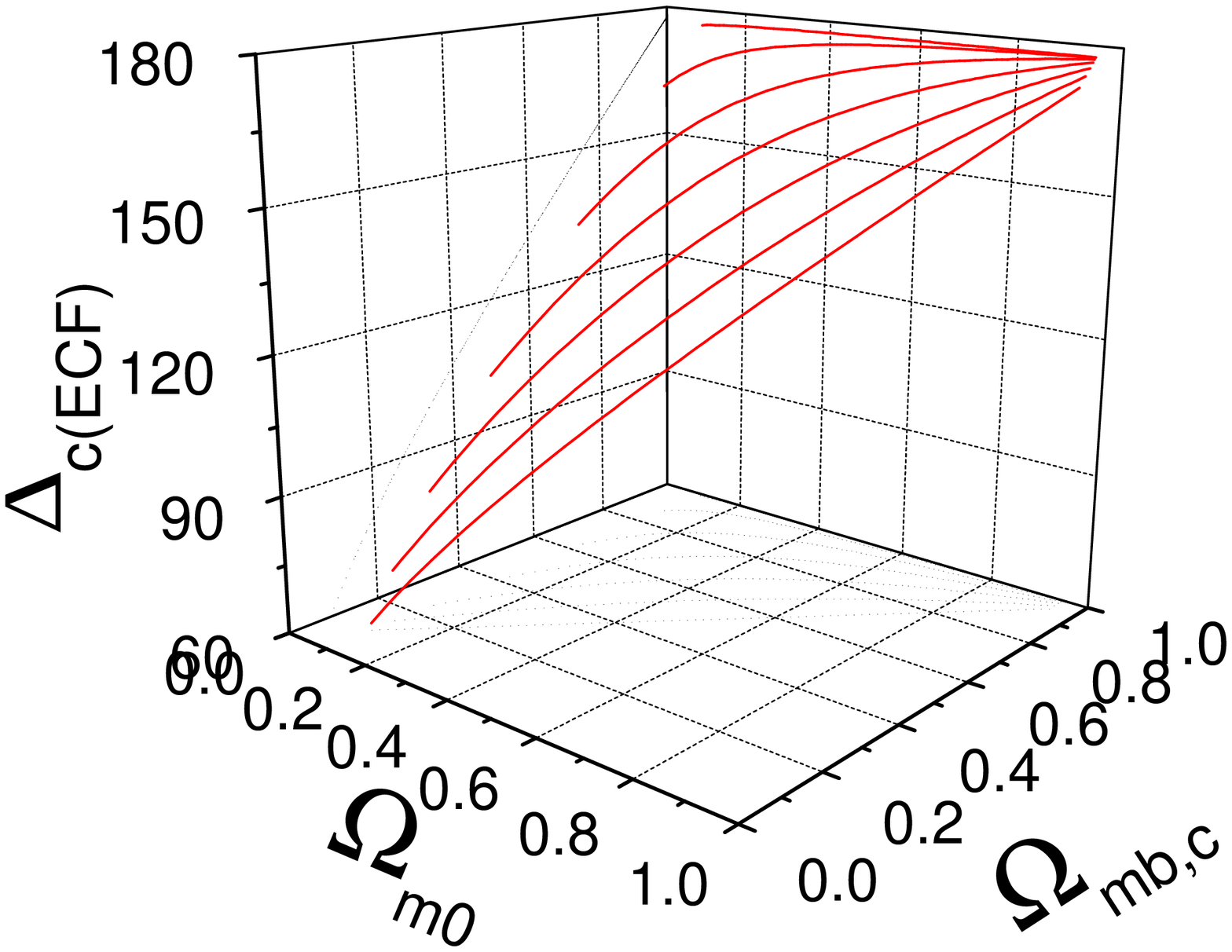}
 \caption{Right panel: $\Delta_{c(ECF)}^{\prime}$s
 dependence on $\Omega_{m0}$ and the collapse point $\Omega_{mb,c}$,
 the same curve has the same collapse epoch.
 Right panel: projection of three space curves from the right
 panel. }\label{ECFD}
 \end{figure}

 To see the physical meaning of
 eq(\ref{NLDeltacAnalytical}) clearly, we plotted
 $\Delta_c^{\prime}$s dependence on $\Omega_{m0}$ and $a_c$
 in FIG.\ref{NLD3D}.
 From the figure, we see that $\Delta_c$ is an increasing function of
 $a_c$, i. e., the latter a region collapses, the more denser
 should it be. This can be understood physically. Because at more later
 times, dark energy's percentage will be more larger. To
 cancel its counter-collapse effects, more denser
 matter is required to assure collapse.
 It is worth noting that
 \cite{ECF} used a different definition of
 $\Delta_{c(ECF)}:=\frac{\rho_{mc,c}}{\rho_{b,totoal,c}}$ (so different
 mass-temperature relations, please comparing eq(2.2) of \cite{ECF}
 and eq(4) of \cite{WangSteinhardt1}), we
 reproduce the results there in FIG.\ref{ECFD}. From the right
 panel of this figure we see that, what was plotted in the lower
 panel of figure.1 of \cite{ECF} is the projection of our results
 on the $\Delta_c$-$\Omega_{mb,c}$ plane.

 \section{Technique Details}
 To calculate parameter $\zeta$ appearing in eqs(\ref{deltacAnalytical}) and
 (\ref{NLDeltacAnalytical}), we write down
 Friedman equations for the back ground cosmology and the space-space
 component of Einstein equation for the radius of the over-dense
 region,
 \beq{}
 (\frac{\dot{a}}{a})^2=\frac{8\pi G}{3}(\rho_{mb}+\rho_\Lambda).
 \label{Friedmann_clusterFormation}
 \eeq
 \beq{}
 \frac{\ddot{r}}{r}=-\frac{4\pi G}{3}
 (3p_{\Lambda}+\rho_\Lambda+\rho_{mc}).
 \label{Einstein_clusterFormation}
 \eeq
 Using the notations introduced in
 eq(\ref{x-y-zeta-nu-definition}),
 eqs(\ref{Einstein_clusterFormation}) and
 (\ref{Friedmann_clusterFormation}) can be rewritten as
 \beq{}
 (\frac{\dot{x}}{x})^2&&\hspace{-3mm}=\frac{8\pi G\rho_{mb,ta}}{3}\frac{1}{x^3\Omega_{mb}(a)}
 \label{Friedmann_clusterFormation2}\\
 \frac{\ddot{y}}{y}&&\hspace{-3mm}=-\frac{4\pi G\rho_{mb,ta}}{3}
 [(1+3w)\frac{1-\Omega_{mb}(a)}{x^3\Omega_{mb}(a)}+\frac{\zeta}{y^3}]
 \label{Einstein_clusterFormation2}
 \eeq
 ,where we preserved the equation of state coefficient $w$ as a
 general parameter. It can even be a function of $a$ or $x$. The
 above two equations can be translated into:
 \beq{}
 &&\hspace{-3mm}\frac{d^2y}{dx^2}
 -\frac{dy}{dx}\frac{1}{2}[\frac{1}{x}+\frac{d\Omega_{mb}}{dx}\frac{1}{\Omega_{mb}}]
 \nonumber\\
 &&\hspace{7mm}+\frac{1}{2}[(1+3w)\frac{y(1-\Omega_{mb})}{x^2}+\frac{\zeta x\Omega_{mb}}{y^2}]
 =0
 \label{zetaDeterminingEq}
 \eeq
 Eq(\ref{zetaDeterminingEq}) is a second ordinary differential equation,
 it contains a free parameter $\zeta$, but satisfies three boundary
 condition:
 \beq{}
 &&[\frac{y}{x}]_{x\rightarrow0}=[\frac{rr^{-1}_{ta}}{aa^{-1}_{ta}}]_{a\rightarrow0}=1_{-}\cdot\zeta^{\frac{1}{3}}
 \nonumber\\
 &&y|_{x=1}=1,\ y^{\prime}|_{x=1}=0.
 \label{BC-zeta-eigenValueEq}
 \eeq
 This is a two point boundary condition problem. It can also
 be looked as an eigen-value problem and solved numerically by the method
 described in \S17.4 of \cite{PressNR}.

 Noting
 \beq{}
 \Omega_{mb}(a)
 =\frac{\Omega_{m0}a^{-3}}{\Omega_{m0}a^{-3}+(1-\Omega_{m0})}
 =\frac{1}{1+\frac{1-\Omega_{mb,ta}}{\Omega_{mb,ta}}x^3},
 \label{OmegaLCDM}
 \eeq
 we see that $\zeta$ solved from eq(\ref{zetaDeterminingEq})
 only depends on $\Omega_{mb,ta}$. So
 we can choose $a_{ta}=1$ and solve eq(\ref{zetaDeterminingEq})
 for different values of $\Omega_{m0}$
 to get the appropriate $\zeta$ and fit
 the result as $\zeta(\Omega_{m0})$. Then change the
 function
 into $\zeta(\Omega_{mb,ta})$ to get $\zeta^\prime$s dependence on
 $\Omega_{m0}$ and $a_c$. We do so by software $Mathematica$ and
 confirmed the results by the method of \cite{PressNR} and get
 \beq{}
 \zeta(\Omega_{mb,ta})=(\frac{3\pi}{4})^2\Omega_{mb}^{-0.7384+0.2451\Omega_{mb}}|_{ta}.
 \label{zetaLCDMold}
 \eeq

 Now the final question we need to answer is to express the
 scale factor of turn around point with that of the collapse
 point.
 Note that, in the unperturbed spherical cases, clusters'
 formation process is symmetrical about the turn around
 time $t_{ta}$, so
 \beq{}
 t_c=2t_{ta}\label{tcAndtta}.
 \eeq
 According to Friedmann
 equation $H^2=H^2_0\Omega_{m0}/a^3\Omega_m$, the time-scale-factor
 relation is
 \beq{}
 t&&\hspace{-3mm}\propto\int_0^{a}da^{\prime}\sqrt{\Omega_m(a^{\prime})a^{\prime}}
 \nonumber\\
 &&=\int_0^{a}da^{\prime}\sqrt{\frac{a^{\prime}}{1+\nu_0 a^{\prime 3}}}
 \nonumber\\
 &&\propto\textrm{ln}[\sqrt{\nu_0 a^3}+\sqrt{\nu_0 a^3+1}].
 \label{time-a-LCDM}
 \eeq
 Substitute eq(\ref{time-a-LCDM}) into (\ref{tcAndtta}) and
 solve it analytically, we get
 \beq{}
 a_{ta}=\left[\frac{\sqrt{1+\nu_0a^3_c}-1}{2\nu_0}\right]^{1/3},
 \label{LCDMtcAndtta}
 \eeq
 where $\nu_0$ is today's ratio of dark-energy/mass density and
 $\nu=\nu_0a^3_{ta}$. \cite{WangSteinhardt1} proposes that when
 eqs(\ref{Friedmann_clusterFormation2}) and (\ref{Einstein_clusterFormation2}) solved,
 using eq(A12) and (A13) of it to calculate $\delta_c$ directly.
 This method has difficulty to get high precision.

 \section{Conclusions}

 We studied the spherical collapse model in the flat $\Lambda$CDM
 cosmology and provided exact and analytical formulaes
 eq(\ref{deltacAnalytical}) and (\ref{NLDeltacAnalytical}) for
 the calculation of the two important parameters
 $\delta_c(\Omega_{m0},a_c)$  and $\Delta_c(\Omega_{m0},a_c)$
 in terms of the media variable $\zeta$ eq(\ref{zetaLCDMold}).
 We reproduced the existing results in the literature, but our method is
 more easier to operate and the result is more simply-looking.
 Our result will be useful in
 the studying of galaxy clusters evolution and the application of Press-Schechter
 theory.

 \section{Acknowledgements}
 We thank very much to profess V. R. Eke for his patience on
 explaining \cite{ECF} to us.


\begin{thebibliography}{10}
\bibitem{Dodelson}Scott Dodelson, 2003, Modern Cosmology, Academic
Press.
\bibitem{PeeblesBigBook}P. J. E. Peebles, 1993, Principles of Physical Cosmology,
Princeton University Press.
\bibitem{PressNR} Press, W. H. et al. 1992, Numerical Recipes,
Cambridge University Press.
\bibitem{PressSchechter} Press W. H., Schechter P., 1974, ApJ,
187, 425
\bibitem{GunnGott}J. E. Gunn, J. R. Gott, APJ 176, 1, 1972.
\bibitem{LaceyCole} C. Lacey and S. Cole, Mon.
Not. Roy. Astro. Soc., 262, 627.
\bibitem{Barrow}J. D. Barrow, Paul Saich,
Mon. Not. Roy. Astro. Soc., 262, 717.
\bibitem{ECF} Vincent R. Eke, Shaun Cole and Carlos S. Frenk, Mon.
Not. Roy. Astro. Soc., 282, 263.
\bibitem{WangSteinhardt1}
Li-Min Wang, Paul J. Steinhardt, Astrophys.J.508:483-490,1998.
astro-ph/9804015
\bibitem{VianaLiddle} Pedro T. P. Viana and Andrew R. Liddle, Mon.
Not. Roy. Astro. Soc., 281, 323.
\bibitem{LokasHoffman}E. L. Lokas and Y. Hoffman,
astro-ph/0011295.
\end{thebibliography}
\end{document}